\documentclass[6pt, onecolumn]{article} 
\usepackage{fullpage}
\usepackage{amsmath}
\usepackage{amssymb}

\usepackage{graphicx}
\usepackage{rotating}

\usepackage{color} 

\usepackage{multicol}

\usepackage[labelfont=bf,labelsep=period,justification=raggedright]{caption}

\makeatletter
\renewcommand{\@biblabel}[1]{\quad#1.}
\makeatother



\makeatletter
\renewcommand{\@biblabel}[1]{\quad#1.}
\makeatother

\date{}

\begin{document}

\begin{flushleft}
\textbf{\Large{Discrete modelling of bacterial conjugation dynamics}}

\bigskip
Angel Go\~ni-Moreno$^{1}$ and Martyn Amos$^{1}$\\

\bf{1} {\small School of Computing, Mathematics and Digital Technology, Manchester Metropolitan University, Manchester M1 5GD, UK.}\\
\medskip
\small{contact: Angel Go\~ni-Moreno: A.Moreno@mmu.ac.uk; Martyn Amos: M.Amos@mmu.ac.uk}

\end{flushleft}

\begin{abstract}
In bacterial populations, cells are able to cooperate in order to yield complex {\it collective} functionalities. Interest in population-level cellular behaviour is increasing, due to both our expanding knowledge of the underlying biological {\it principles}, and the growing range of possible {\it applications} for engineered microbial consortia. Researchers in the field of synthetic biology - the application of engineering principles to living systems - have, for example, recently shown how useful decision-making circuits may be {\it distributed} across a bacterial population. The ability of cells to interact through small signalling molecules (a mechanism known as {\it quorum sensing}) is the basis for the majority of existing engineered systems. However, horizontal gene transfer (or {\it conjugation}) offers the possibility of cells exchanging messages (using DNA) that are much more information-rich. The potential of engineering this conjugation mechanism to suit specific goals will guide future developments in this area. Motivated by a lack of computational models for examining the specific dynamics of conjugation, we present a simulation framework for its further study. We present an agent-based model for conjugation dynamics, with realistic handling of physical forces. Our framework combines the management of intercellular interactions together with simulation of intracellular genetic networks, to provide a general-purpose platform. We validate our simulations against existing experimental data,  and then demonstrate how the emergent mixing patterns of multi-strain populations can affect conjugation dynamics.  Our model of conjugation, based on a probability distribution, may be easily tuned to correspond to the behaviour of different cell types. Simulation code and movies are available at http://code.google.com/p/discus/.
\end{abstract}

\begin{multicols}{2}

\section{Introduction}
Researchers in the emerging field of synthetic biology \cite{Heinemann2006} have demonstrated the successful construction of devices based on populations of engineered microbes \cite{Basu2005,Sole2012,Tamsir2011}. Recent work has focussed attention on the combination of single-cell intracellular devices \cite{Gardner2000, Auslander2012} with intercellular engineering, in order to build increasingly complex systems. 

To date, most work on engineered cell-cell communication has focussed on quorum-sensing (QS) \cite{Atkinson2009}, which may be thought of as a communication protocol to facilitate inter-bacterial communication via the generation and receiving of small signal molecules. However, recent studies on DNA messaging \cite{Endy2012} highlight the importance and utility of transferring whole sets of DNA molecules from one cell (the so-called donor) to another (the recipient). Bacterial {\it conjugation} is a cell-to-cell communication mechanism \cite{Fernando2010, Fernando2012} that enables such transfers to occur.

In this paper we present a simulation platform that realistically simulates (in a modular fashion) both intracellular genetic networks and intercellular communication via conjugation. To our knowledge, this is the first such platform to offer both of these facilities. We first review previous work on cell simulation, before presenting the details of our model. We validate it against previous experimental work, and then discuss possible applications of our method.

\section{Previous work}

The rapid development of bacterial-based devices is accompanied by a need for computational simulations and mathematical modelling to facilitate the characterisation and design of such systems. A number of of platforms and methods are available for this purpose. Agent-based models (AbMs) are widely used \cite{BSim2012}, and were first used to study microbial growth in {\it BacSim} \cite{Kreft1998}. Continuous models have also been proposed \cite{Melke2010}, and recent developments make use of hardware optimisation, by using GPUs (Graphics Processing Units) in order to scale up the number of cells simulated \cite{Haseloff2012}.

Because of the complexity of the system under study, several computational platforms focus on either specific cellular behaviours (e.g., bacterial chemotaxis \cite{Emonet2005}, morphogenesis of dense tissue like systems \cite{Izaguirre2004}) , or on specific organisms (e.g., {\it Myxococcus xanthus} \cite{Holmes2010}).  Platforms that incorporate cell-cell communication generally focus their attention on quorum-sensing. Simulations of conjugation do exist, but these consider cells as abstracted {\it circular objects} \cite{Krone2007,Seoane2011b}. We demonstrate in this paper how a consideration of the {\it shape} of cells is an essential feature for understanding the conjugation behaviour of the population. We now describe our model for bacterial growth, in which conjugation is handled explicitly. \\ \\ 

\section{Methods}
\label{Methods}

We use an {\it individual-based modelling} approach \cite{Lardon2011} 
to the study of conjugation dynamics. This models each cell as an individual, mobile entity, each of which is subject to physical forces arising from contact with other cells and the environment (e.g., surfaces). Each cell has a number of different {\it attributes}, listed in Table ~\ref{tab:variables2}, which correspond to various physiological states and characteristics.

\begin{table*}[htbp]
\centering
\caption{Cell attributes.}
{\begin{tabular}{l l l}
\hline
Attribute & type & Definition \\
\hline
\texttt{shape} & {\it pymunk}.Shape & Shape of the cell\\
\texttt{program} & [m$_{0}$ $\ldots$ m$_{i}$] & List of the {\it i} regulatory network molecules (m)\\
\texttt{elongation} & [int,int] & Elongation values (one per cell pole) \\
\texttt{position} & [x,y] & Coordinates of centre point, {\it x} and {\it y}\\
\texttt{speed} & float & Velocity\\
\texttt{conjugating} & Boolean & Conjugation state\\
\texttt{plasmid} & Boolean & Program state (present/not present)\\
\texttt{role} & int & Donor (0), recipient (1) or transconjugant (2)\\
\texttt{partner} & int & Role of plasmid transfer cell \\
 \hline
\label{tab:variables2}
\end{tabular}}{}
\end{table*}

\begin{figure*}[!tpb]
\centerline{\includegraphics[width=60mm]{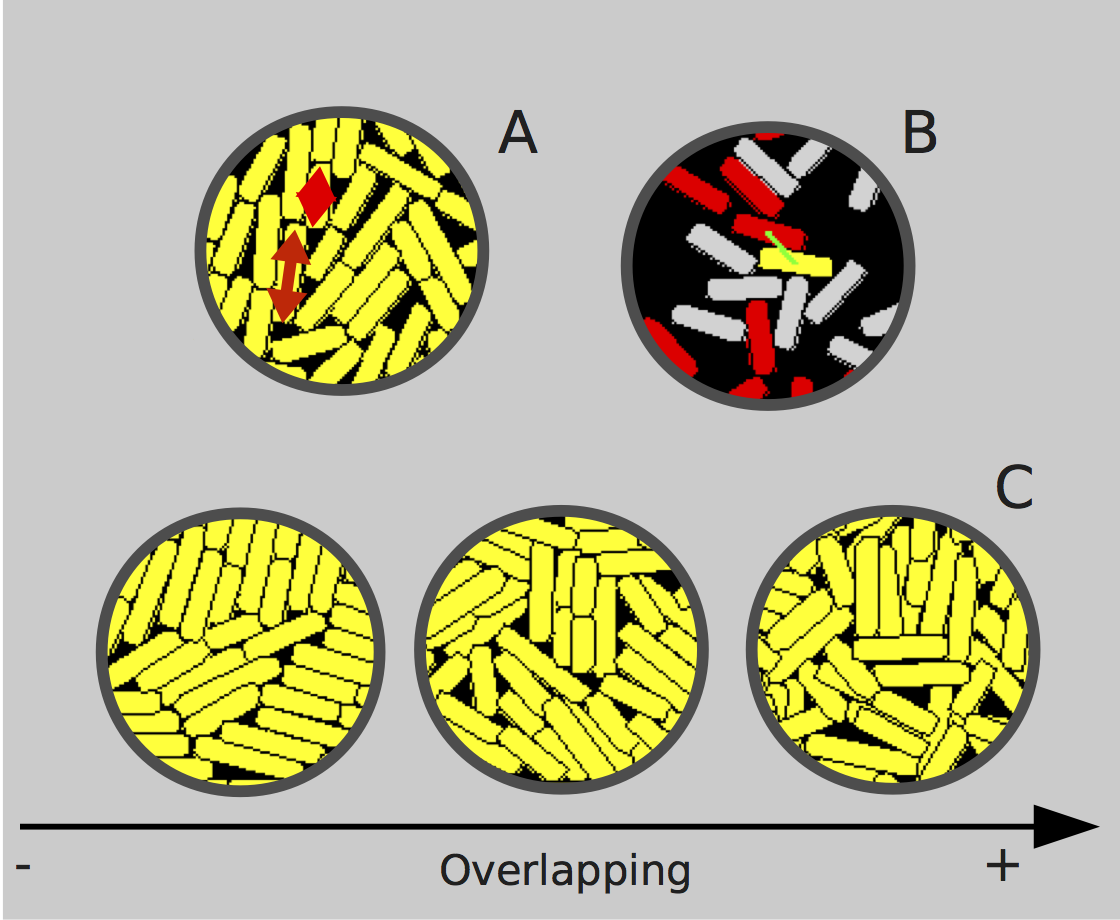}}
\caption{Cell behaviours at low scale. (A): Different cell length due to asynchronous growth (pressure dependent). Two cells marked with red arrows. (B): A donor cell (red) starts the conjugation process with a recipient (grey) which turns into transconjugant (yellow). The pilus (green) is an elastic spring that links the two cells until the process is finished. (C): Different overlapping levels within the cells of a population.}
\label{fig:intro}
\end{figure*}

Bacteria are modelled as rod-shaped cells with a constant radius (parameter \texttt{width} in Table \ref{tab:variables}). Elongation processes occur along the longitudinal axis, which has a minimum dimension of \texttt{length}, and division takes place whenever a the cell measures 2*\texttt{length}. The division of a cell into two new daughter cells is also controlled by \texttt{max\_overlap}, which monitors the physical {\it pressure} affecting each cell; if the pressure exceeds this parameter value, the cell delays its growth and division. Thus, a cell with pressure grows slower than without it. In Figure \ref{fig:intro}A we see an snapshot of a population with different cell lengths, due to the pressure-dependent behaviour. The global parameter \texttt{growth\_speed} (Table ~\ref{tab:variables}) also helps us simulate cell flexibility in a realistic fashion. This parameter defines a ``cut off" value for the number of iterations in which the physics engine must resolve {\it all} the current forces and collisions. Thus, smaller values will cause the solver to be effectively ``overloaded", and some collisions may, as a result, be partially undetected. This means that cells behave as flexible shapes, which gives the simulation a more realistic performance. In Figure \ref{fig:intro}C we show how changes in \texttt{growth\_speed} affect the simulation, using bigger (left) to smaller (right) values.

Horizontal genetic transfer (or conjugation) is modelled using an {\it elastic spring} to connect donor and recipient cells. Parameter \texttt{c\_time} defines the duration of that linkage, which determines the time in which the DNA is transferred. The springs are constantly monitored to ensure that they physically connect both cells during conjugation. Importantly, during conjugation, the resolution of collisions involving relevant cells considers the forces produced by the spring connection, in order to calculate the final movement of the bacteria. By coupling cells in this way, we obtain realistic population-level physical patterns that emerge as a result of large numbers of conjugation events. Figure \ref{fig:intro}B shows this process, with a donor cell (red) and a recipient cell (grey) which becomes a transconjugant (yellow). A transconjugant cell is one that was initially a recipient, but which has been conjugated during its lifetime. Thus, it already has the DNA information transferred by the donor.

This agent-based algorithm has an iteration-driven structure, where - after initialisation of the main global parameters - it repeatedly performs the following steps for each cell: \\ \\

\begin{enumerate}
\item Update springs (position and timing).
\item Perform cell division (if cell is ready).
\item Elongate cell (every \texttt{growth\_speed} steps).
\item Handle conjugation.
\item Update physical position.
\end{enumerate}

Conjugation decisions (step 4) made by cells are driven by three sequential steps:

\begin{enumerate}
\item The cell {\it decides}, following a probability distribution, whether or not to conjugate (one trial per iteration).
\item If conjugating, randomly select a mate from surrounding bacteria (if present).
\item If valid mate is found, effect conjugation transfer.
\end{enumerate}

The discrete probability distribution used for the conjugation process is $C(N, p, \texttt{c\_time})$, where $N$ is the number of trials in a cell lifetime (\texttt{width} * \texttt{length}), $p$ is the success probability in each trial (with p $\in$ [0$\ldots$1]) and \texttt{c\_time} is the time interval during $p = 0.0$ (i.e., when the cell is already conjugating). As stated in \cite{Seoane2011a}, $p$ can vary, depending on whether the cell is a donor (\texttt{p\_d}), a transconjugant that received the DNA message from a donor (\texttt{p\_t1}), or a transconjugant that received the DNA from another transconjugant (\texttt{p\_t2}).

\begin{sidewaystable*}
\centering
\caption{Global simulation parameters.}
{\begin{tabular}{ l l |  l l}
\hline
Parameter & Definition & Parameter & Definition \\
\hline
\texttt{screenview} & Size of the simulated {\it world} & \texttt{network\_steps} & Number of steps of the ODEs per \texttt{Gt}\\
\texttt{max\_overlap} & Pressure tolerance of cells  & \texttt{number\_donors} & Initial number of donor cells\\
\texttt{width} & Width of each cell (lattice squares) & \texttt{number\_recipients} & Initial number of recipient cells \\
\texttt{length} & Length of each cell (lattice squares) & \texttt{spring\_rest\_length} & Natural sprint expansion/contraction\\
\texttt{growth\_speed} & Iterations between elongation processes & \texttt{spring\_stiffness} & The tensile modulus of the spring\\
\texttt{Gt} & Doubling time of the simulated cells (iterations) & \texttt{spring\_damping} & The amount of viscous damping to apply\\
\texttt{real\_Gt} & Real doubling time of the studied cells (minutes) & \texttt{cell\_infancy} & Time lag (percentage)\\
\texttt{p\_d} & Probability of conjugation event (donors) & \texttt{pymunk\_steps} & Update the space for the given time step\\
\texttt{p\_t1} & Probability of conjugation event (transconj.1) & \texttt{pymunk\_clock\_ticks} & Frame frequency (FPS - frames per second\\
\texttt{p\_t2} & Probability of conjugation event (transconj.2) & \texttt{bac\_mass}& Mass of the cell (for calculating the moment)\\
\texttt{c\_time} & Duration of the conjugation process & \texttt{bac\_friction} & Friction coefficient (Coulomb friction model)\\
\hline
\label{tab:variables}
\end{tabular}}{}

\end{sidewaystable*}

Intracellular circuitry is modelled separately, and then {\it introduced} into each cell by storing the state of the circuit in an attribute of the cell (\texttt{program}). Thus, there are effectively as many copies of the circuit as cells in the simulation. This circuit simulation is implemented in a modular fashion, so that the internal cellular ``program" may be easily replaced with any other. In this paper we demonstrate the principle using a two-component genetic oscillator as the DNA message that is exchanged through conjugation. The ordinary differential equations (ODEs) for this circuit are: 

\begin{equation}
\frac{dx}{dt} = \Delta \left ( \beta \frac{1 + \alpha x^2}{1 + x^2 + \sigma y^2 } - x\right ) \label{eq1}
\end{equation}
\begin{equation}
\frac{dy}{dt} = \Delta \gamma \frac{1 + \alpha x^2}{1 + x^2} - y \label{eq2}
\end{equation}

\noindent which are detailed in \cite{PoyatosPLoS}, as well as the meaning and value of the parameters (we use the same values in the code provided). We have also recently used our software platform to investigate the spatial behaviour of a {\it reconfigurable} genetic logic circuit \cite{Reconfigurable}, which demonstrates how it may easily be modified to accommodate different sets of equations.
The actions controlling the growth rates of cells occur on a longer time scale than the integration steps that control molecular reactions (as equations \ref{eq1} and \ref{eq2}). In order to ensure synchronisation, the parameter \texttt{network\_steps} defines the number of integration steps of the ODEs that run per \texttt{Gt}. Thus, a number of \texttt{network\_steps}/\texttt{Gt} integration steps will update the attribute \texttt{network} of each cell every iteration.

Other important physical parameters listed in Table ~\ref{tab:variables} are \\ \texttt{spring\_rest\_length}, \texttt{spring\_stiffness} and \texttt{spring\_damping}; these are three parameters to model the material and behaviour of the bacterial pilus (i.e. the spring) during conjugation. Parameter \texttt{cell\_infancy} is a delay period, during which a cell is considered to be too young to conjugate (as observed experimentally \cite{Seoane2011a}). Parameters \texttt{pymunk\_steps} and \texttt{pymunk\_clock\_ticks} are used by the physics engine to update the world, and may be adjusted  by the user in order to alter the performance of the simulation (machine dependent). Parameters \texttt{bac\_mass} and \texttt{bac\_friction} play a role in collision handling. 

Our platform is writting in {\it Python}, and makes use of the physics engine {\it pymunk} (www.pymunk.org) as a wrapper for the 2D physics library Chipmunk, which is written in C (www.chipmunk-physics.net/). As cells are represented as semi-rigid bodies in a 2D lattice, pymunk handles the physical environment on our behalf.  For monitoring purposes, parameters \texttt{Gt} and \texttt{real\_Gt} allow us to stablise the relation between iterations and clock minutes: $minute = \texttt{Gt}/\texttt{real\_Gt}$ (units: iterations). The platform, which we call DiSCUS (Discrete Simulation of Conjugation Using Springs) is available at the project repository at http://code.google.com/p/discus/.


\begin{figure*}[!tpb]
\centerline{\includegraphics[width=150mm]{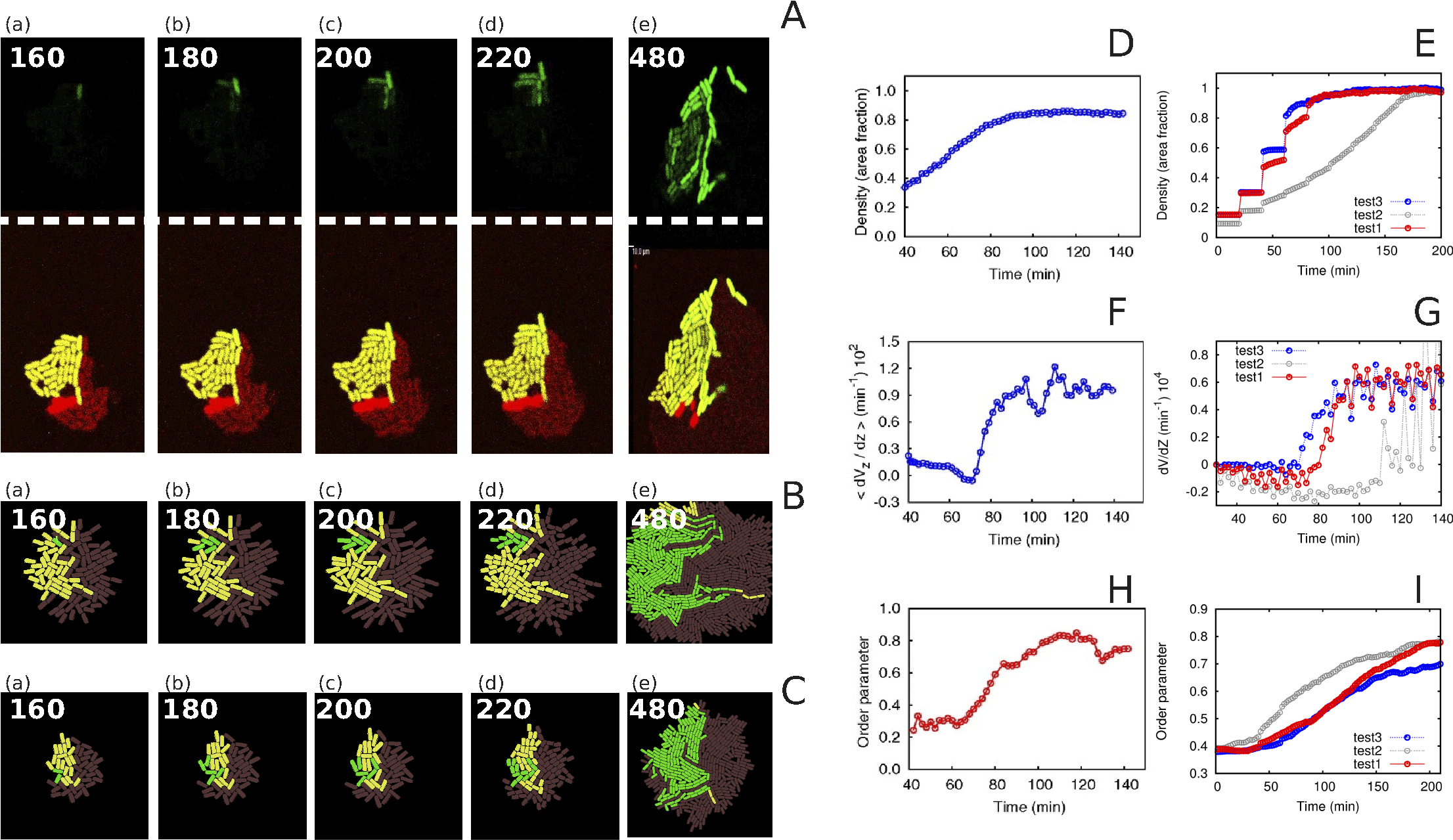}}
\caption{Validation of cell movement and conjugation dynamics using real data. (A): Figure extracted from \cite{Seoane2011a} where a colony of {\it Pseudomonas putida} is divided into dark red donor cells (DsRed), yellow recipient cells (YFP) and transconjugants, expressing both yellow and green light (YFP and GFP). The upper row shows the transconjugant signal, and the bottom row shows the whole community. (B and C): Simulation results. Two simulations of similar colonies are recorded over exactly the same time intervals (min). The colours of the cells match the colours observed in (A). Graphs (D), (F) and( H) are extracted from \cite{Volfson2008},  and show experimental results of {\it Escherichia Coli} growth regarding density, velocity and ordering (respectively). Graphs (E), (G) and (I) correspond to simulations in similar conditions to \cite{Volfson2008}, for the same parameters (density, velocity and ordering respectively). Tests 1, 2 and 3 in graphs correspond to different spatial distribution of cells inside the microfluidic chanel (details in text).}
\label{fig:validation}
\end{figure*}

\section{Results}
\label{Results}

We now describe the results of experiments to validate our conjugation model, using four sets of simulations. As we aim to understand the behaviour of cells in small-scale two-dimensional populations (as occur in microfluidic environments), we avoid the sorts of extreme overlapping situations shown in Figure \ref{fig:intro}C(right). We first validate individual conjugation dynamics; then we validate the biomechanical properties of the simulation; the third set of experiments concerns the transfer of the two-component oscillator, and the final set of experiments study the effects of mixing on conjugation dynamics. We now describe in detail the results of each set of experiments.

\subsection{Conjugation dynamics}

\begin{figure*}[!tpb]
\centerline{\includegraphics[width=158mm]{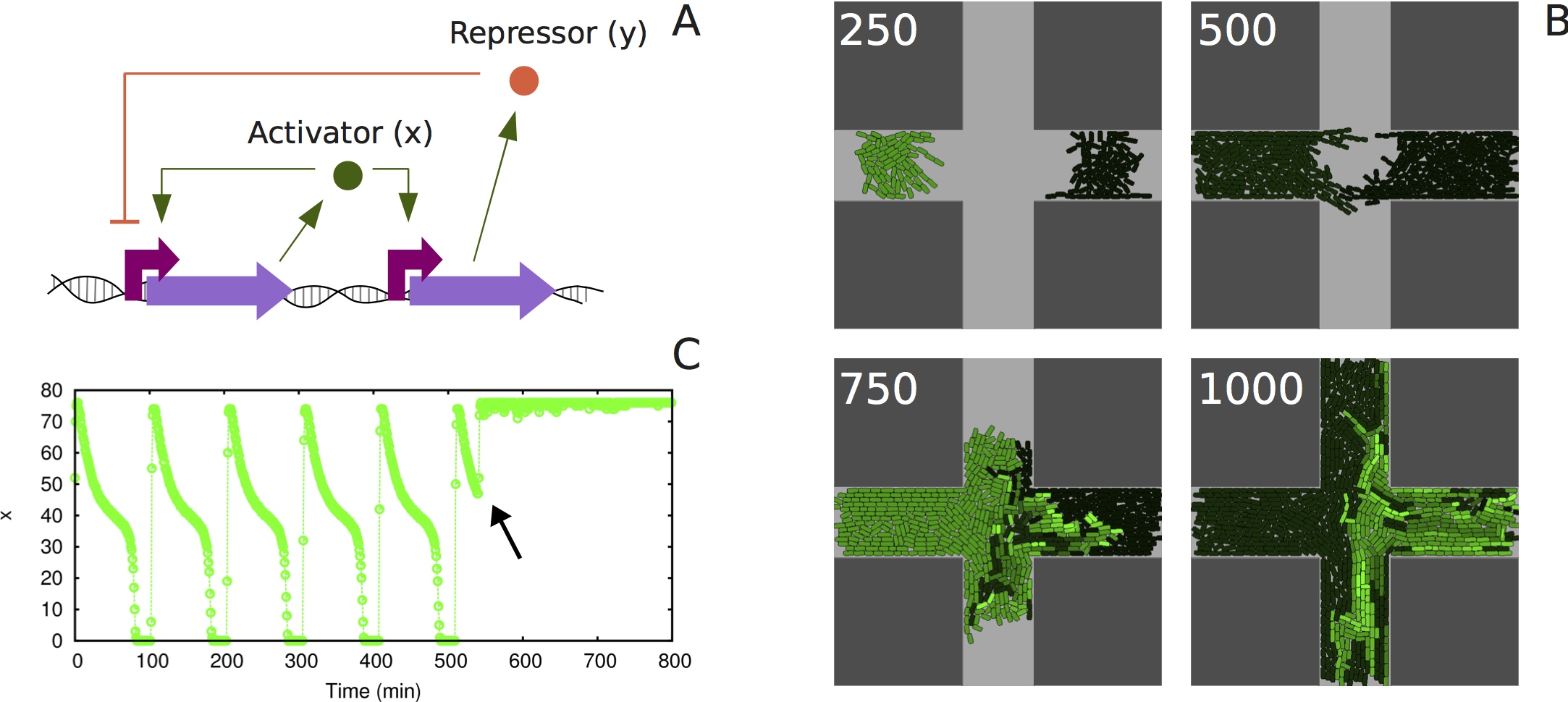}}
\caption{Horizontal transfer of a two-component genetic oscillator. A: Transcriptional-level design. The activator (green) acts on itself and on the repressor (red) by inducing the transcription of both. The represor acts on the activator by repressing its transcription. As a result, molecule {\it x} (as well as {\it y} oscillates in time. B: Cells growing in a cross-shaped channel. At 250 (minutes) we clearly see the position of both donor (left hand, with the oscillator inside) and recipient (right hand, empty) strains. The intensity of green colour denotes the amount of molecule {\it x} inside cells in that specific time. Through conjugation, the oscillator is {\it copied} to the initial recipient strain (t = 1000 min). C: Using the same experiment as in B, measurement over time of the maximum level of molecule {\it x} in a single cell. Black arrow highlights the point in time when conjugation starts (t $\simeq$ 550 min).}
\label{fig:oscillator}
\end{figure*}

The objective of the first set of experiments is to validate the software in terms of {\it conjugation dynamics}. For that purpose, we first focus on conjugation, using images of a {\it Pseudomonas putida} population (Figure \ref{fig:validation}A) extracted with permission from \cite{Seoane2011a}. These show donor cells (dark red) growing in contact with recipients (yellow). The DNA information they share after conjugation makes the transconjugant cells display GFP (green fluorescent protein). We adjusted the parameters of our simulations until the behaviour matched the images of real cells (two simulations shown: Figures \ref{fig:validation}B and \ref{fig:validation}C), in terms of both time-series behaviour and the type of physical pattern displayed. It is important to note that the differential probabilities of conjugation of donors and transconjugants (higher in the latter) causes directional spreading of the DNA information. After the first transconjugant appears (160 minutes), the newly-formed transconjugants appear  -most probably - in the immediate neighbourhood.  The final parameter values used to reproduce this experiment are: \texttt{width}=5, \texttt{length}=15, \texttt{growth\_speed}=30, \texttt{p\_d}=0.001, \texttt{p\_t1}=0.02, \texttt{p\_t2}=0.05 and \texttt{c\_time}=450 (the rest of the parameters are as defined in the DiSCUS distribution). Movie {\it DemoConjugation1} (found in the project repository) shows a simulation of a similar experiment where the transconjugants do not act as new donors.

\subsection{Biomechanical properties}

The second set of validation experiments focuses on {\it biomechanical movement}. We use data from \cite{Volfson2008}, which describe an {\it Escherichia coli} colony growing in a microfluidic channel (30 * 50 * 1 $\mu$m$^3$) (Figures \ref{fig:validation}D, \ref{fig:validation}F and \ref{fig:validation}H). Using the same parameter setup (\texttt{width}=5, \texttt{length}=24, \texttt{growth\_speed}=30) we highlight how different initial positioning of cells inside the channel can affect the final result ({\it test1}, with more cells observed in the centre than at the edges; {\it test2} with all cells initially in the centre; {\it test3} with all cells homogeneously spread along the channel). Density graphs (Figures \ref{fig:validation}D and \ref{fig:validation}E) show the increasing curve as the channel becomes more populated (results vary depending on which area is considered for monitoring). Velocity gradients (Figures \ref{fig:validation}F and \ref{fig:validation}G) depict the differential velocity across the longitudinal axis of the channel with respect to the centre (we see negative values when the cells in the centre move faster than the rest). The difference in the {\it y} axis is due to our considering different spacial intervals in the velocity gradient calculation. Ordering graphs (Figures \ref{fig:validation}H and \ref{fig:validation}I) are based on calculating the cosine of a cell's angle with respect to the longitudinal axis of the channel (e.g. angle 0, cos(0)=1, completely aligned). As time increases, we see that the cells tend to align themselves. 

\subsection{Internal cell ``program"}

Figure \ref{fig:oscillator} shows the results of the next set of experiments, aiming at studying the horizontal transfer of a {\it two-component oscillator}. The transcriptional-level design of the circuit is shown in Figure \ref{fig:oscillator}(A), which causes the molecular concentration of a repressor (y) and an activator (x) to oscillate over time. Each molecule is produced by a gene when its upstream promoter is activated. The activator can induce its own production at the same time as inducing the production of the repressor, which in turns inhibits the production of the activator. In Figure \ref{fig:oscillator}B we place (250 minutes) an initial donor colony (with the two-component oscillator inside) on the left-hand side of a cross-shaped channel, while a recipient colony is placed on the right.

As the equations for the oscillator have no stochasticity, every cell of the donor strain shows exactly the same state of the circuit as every other cell. At the beginning (250 minutes), these cells show a green colour (corresponding to the molecular concentration of x) which is switched off during the time intervals in which the repressor is {\it on} (see time profile of Figure \ref{fig:oscillator}C). When conjugation starts (at around t $\simeq$ 550), the newly formed transconjugant cells are given the circuit but, importantly, they do not share the state of the circuit of the cells from which they receive the message. During the DNA transfer, it is only the plasmid (circuit {\it carrier}) that is copied into the recipient;  therefore, both molecular concentrations are null, and the circuit begins its functioning from the initial stage. That is why we clearly observe different green intensities within the community. This asynchronous behaviour happens only in the transconjugants, while the circuits inside the donors always run synchronously (due to deterministic equations).

A time profile of the previous experiment is shown in Figure \ref{fig:oscillator}C, where the maximum level of activator (x) concentration in a single cell (compared with the whole population) is recorded over time. Before conjugation starts, all cells in the consortia display perfect synchrony. After conjugation (shown with an arrow on the graph) there is always a cell with the maximum level of activator, which demonstrates high asynchrony. All parameter values regarding cell dimensions or conjugation probabilities are the same as in Figures \ref{fig:validation}B and \ref{fig:validation}C. Parameters relevant to the oscillator are: \texttt{network\_steps}=18 and \texttt{Gt}=450. Movie {\it DemoConjutagion2} (found in the project repository) shows this experiment and Movie {\it DemoDynamics1} shows donor cells growing with a stochastic version of the oscillator.

\subsection{Effects of mixing}

\begin{figure*}[!tpb]
\centerline{\includegraphics[width=150mm]{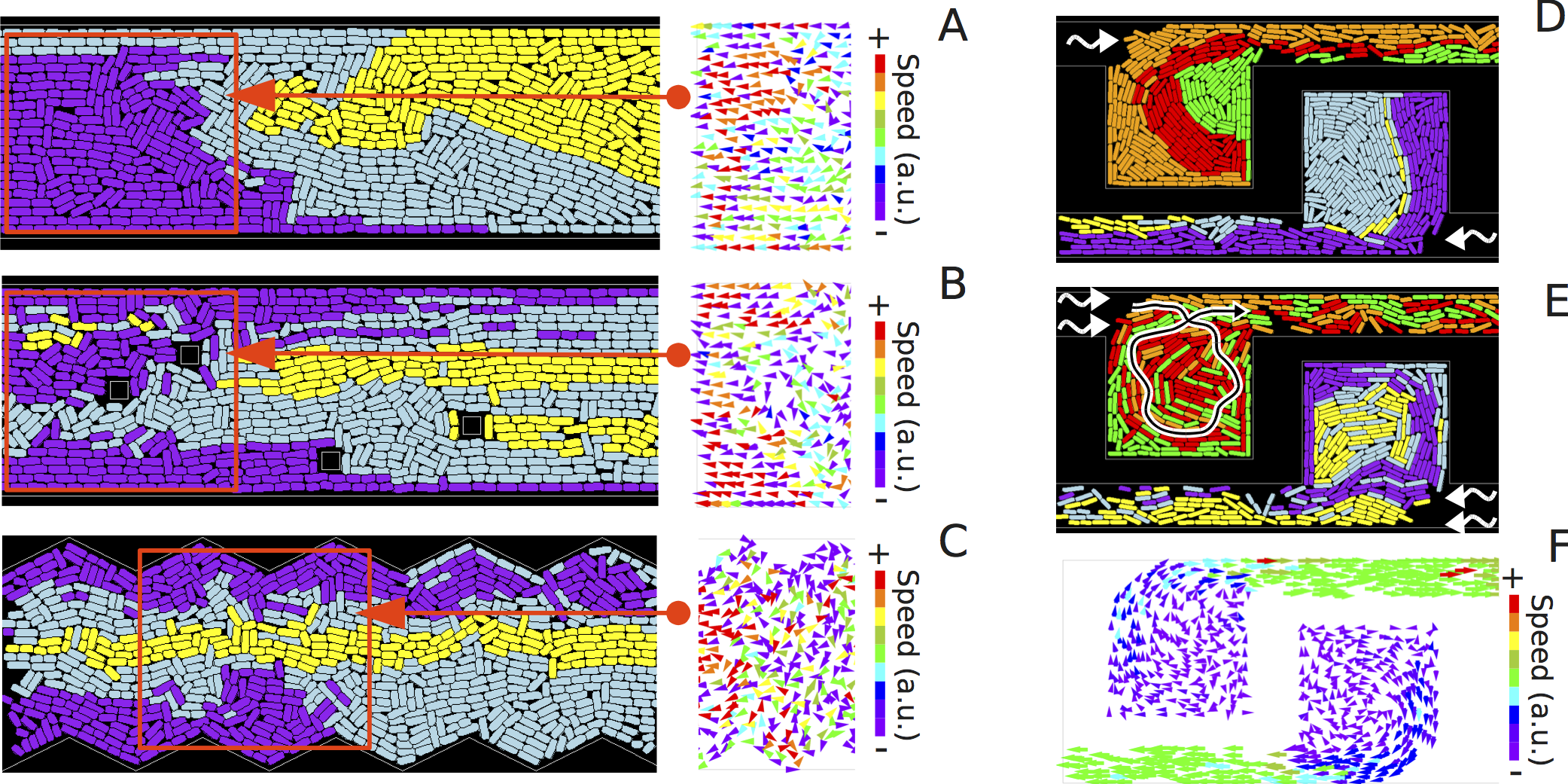}}
\caption{Dynamical mixing of bacterial strains under different conditions. (A): Three strains (purple, yellow and blue) growing in a longitudinal pipe. Detail (on the right) shows the vector field corresponding to the red square (on the left) where the arrows display both directionality and velocity of every cell. (B): Similar to (A), but with four {\it columns} placed in the middle of the pipe. (C): Similar to A, but with {\it zig-zag} borders along the pipe. In (A), (B) and (C), the speed in vector fields is measured in arbitrary units (a.u.). (D): Six strains (three per trap) grow in two square traps on one side of the main longitudinal channel. The flow in the channels follows the direction of the arrows. (E): Similar to (D) but with a much stronger flow in channels, causing turbulence in traps (long circled arrow). (F): Vector field of experiment (E). }
\label{fig:mixing}
\end{figure*}

\begin{figure*}[!tpb]
\centerline{\includegraphics[width=60mm]{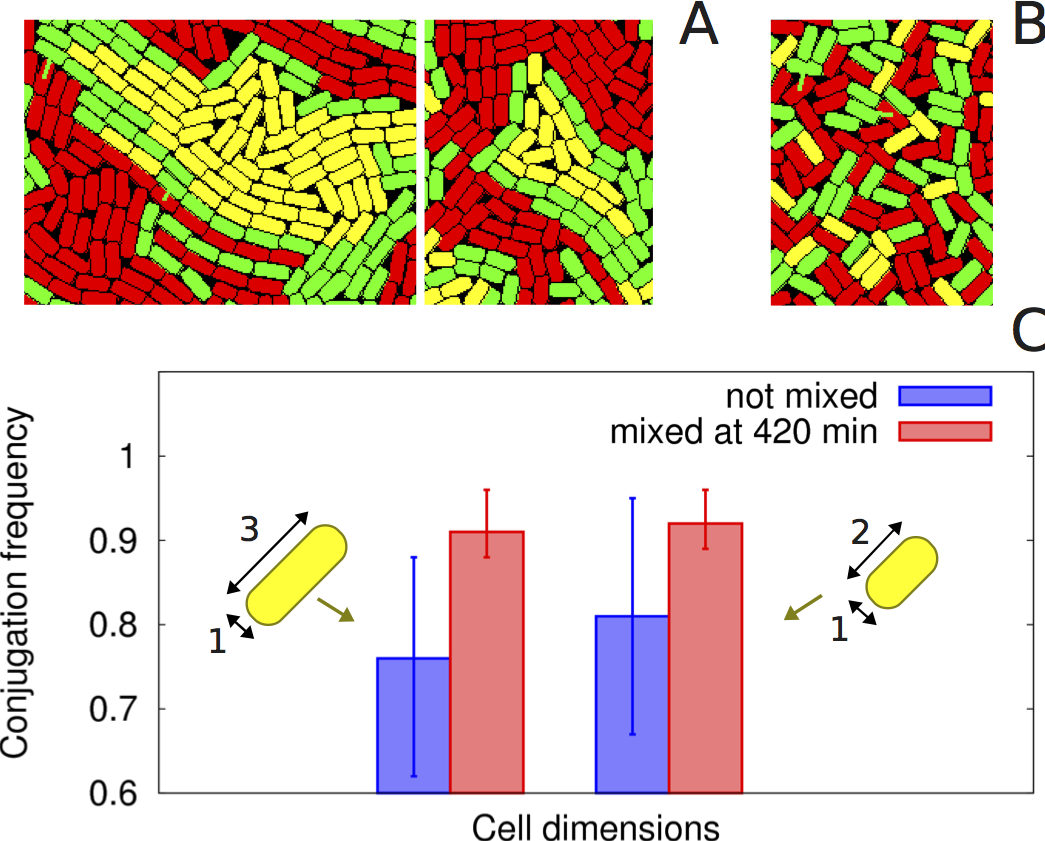}}
\caption{Effects of manual mixing on conjugation frequency. (A): Recipient-trapping behaviour of a population with donors (red), transconjugants (green) and recipients (yellow). Two snapshots depict clearly-observed clusters. (B): Population after random mixing,  where the clusters are automatically dissolved. (C): Graph showing conjugation frequencies (Y = T/(R + T)) of 560-minute experiments (ratio D/R = 50\%). Blue bars represent Y on an untouched population, while red bars represent Y when the population is mixed at 420 minutes. The two sets of bars correspond to experiments with different cell dimensions (1x3 -left- and 1x2 -right). Error bars show variation across 15 experiments of each class.}
\label{fig:frequency}
\end{figure*}

Conjugation behaviour within a population may be altered in different ways to achieve different behaviours, depending on the desired application. For example, in the previous experiments described in this paper, transconjugants are unable to act as recipients (simulating a {\it radical} entry exclusion \cite{Mapi2008}). That is to say, they will not receive more plasmids (genetic circuits) from either donors or transconjugants. Furthermore, we may also engineer the transconjugants to stop acting as {\it new donors} \cite{Fernando2012}, so that only the original donors have the ability to transfer the DNA message. {\it Mixing} of the cell population becomes essential in this last scenario, in order to ensure maximal contact between donors and recipients. In Figure \ref{fig:oscillator}B we see how, at the end of the experiment (1000 minutes), donors and the transconjugants cover different areas of the channel (left and right respectively), without being mixed.

We now study the autonomous mixing behaviour of cells under different environmental conditions, with the third sets of experiments (Figure \ref{fig:mixing}). Firstly we investigate how morphological changes in a longitudinal microfluidic channel can affect the patterns being formed by the consortia and its mixing. Figure \ref{fig:mixing}A shows three bacterial strains (each shown in a different colour) growing in a channel from different starting points. As we can see, their mixing is highly improbable. The main reason for this is the velocity and directionality of the cells. As the cells are washed out at the edges of the channel, all of them {\it travel} (they only have passive movement while being pushed) at variable speed from centre to left or from centre to right. This causes the cells to have the same direction (see vector field), which in turn makes mixing more difficult. In Figures \ref{fig:mixing}B and \ref{fig:mixing}C we show the result of altering the morphology of the channels, by adding columns and zig-zag walls, respectively. As a result, the cells show different directionality (see vector fields) and the strains have a higher probability of being mixed. In both experiments (unlike \ref{fig:mixing}(A)), the three strains are in contact at some point. If the experimental application relied on the conjugation of purple and yellow cell pairs, for example, we can see that these physical changes in the channel would be essential.

Another way to intensify the mixing of strains in a microfluidic trap is to change the main channel {\it flow strength},  with the objective of creating turbulence inside the trap. Figures \ref{fig:mixing}D and \ref{fig:mixing}E show a three-strain population growing in a trap (identical initial positioning of cells in both) with the only difference being that the strong flow in the main channel (white arrows) of \ref{fig:mixing}E creates turbulence (in the direction of the circled arrow). Two different colonies are simulated in each experiment, inside both symmetrical and independent traps. We see how turbulence helps the cells to get mixed, thanks to the constant change of direction they display (see vector field in Figure \ref{fig:mixing}F). Furthermore, we can avoid {\it missing} one strain, as happens with the yellow cells in \ref{fig:mixing}D (see Movie {\it DemoDynamics2} -found in the project repository-), where another run of this experiment is shown).

Investigations of how manual mixing can affect conjugation frequencies are described in in \cite{Fernando2012}, using an {\it Escherichia coli} population. We now reproduce those results using our software, and give valuable insight into the reasons for that behaviour: the {\it isolation} of the recipients. For that purpose (Figure \ref{fig:frequency}) we grow a population of donors (D, red) and recipients (R, yellow) in which the ratio D/R is 50\% and the transconjungants (T, green) are unable to act as new donors. The frequency of conjugation, Y, is measured as Y = T/(R + T). The graph in Figure \ref{fig:frequency}C shows the frequency after 560 minutes of {\it untouched} populations (not mixed, blue bars) and populations that have been {\it manually mixed} at 420 minutes (red bars). The difference that the mixing produces is based on the isolation of the recipients in untouched populations. Figure \ref{fig:frequency}A shows two different occasions in which clusters of recipients are formed, where the transconjugants do not allow donors to reach new possible mates. After the population is completely ``shuffled" (\ref{fig:frequency}B), the clusters are dissolved, and new pairs of donor-recipient can arise in the new topology.

An interesting result from Figure \ref{fig:frequency}C is the fact that the smaller the size of the cell, the higher results we observe for conjugation frequencies. This may be due to the fact that smaller cells are able to slip through physical gaps, and the biomechanical ordering of the population becomes more ``fuzzy". This underlines the importance of considering the physical shape of cells, since circle-shaped cells would not give valid results.

\section{Discussion}

The conjugation model presented here is the first agent-based model to explicitly simulates the conjugation process with growing rod-shaped cells. Full validation against real data is performed, which shows the capacity of the software to reproduce observed behaviour.  In addition, the mixing study offers valuable insights into the design of multi-strain populations. The software also allows for genetic {\it programs} to be {\it installed} inside cells; the potential for horizontal gene transfer to recreate distributed information processing within a microbial consortium is of significant interest in synthetic biology, and the software presented will aid the design and testing of systems before their {\it wet-lab} implementation.

\bibliography{template}

\end{multicols}

\end{document}